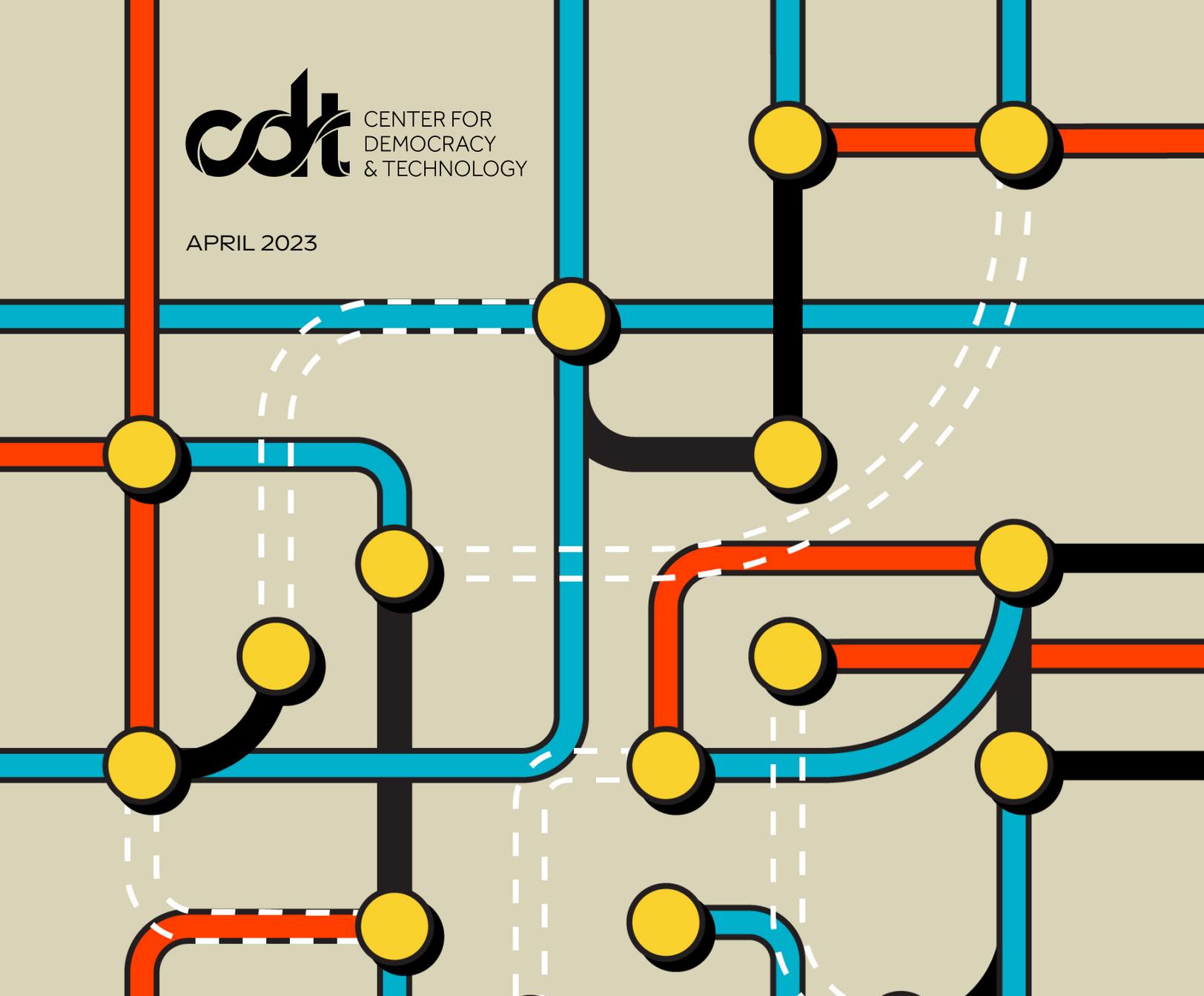

CENTER FOR
DEMOCRACY
& TECHNOLOGY

APRIL 2023

# Slicing the Network:

## Maintaining Neutrality, Protecting Privacy, and Promoting Competition

A technical and policy overview with recommendations for operators and regulators

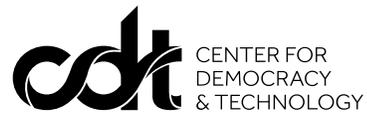

The Center for Democracy & Technology (CDT) is a 28-year-old 501(c)3 nonpartisan nonprofit organization working to promote democratic values by shaping technology policy and architecture, with a focus on equity and justice. The organization is headquartered in Washington, D.C. and has a Europe Office in Brussels, Belgium.

# Slicing the Network:

## Maintaining Neutrality, Protecting Privacy, and Promoting Competition

A technical and policy overview with recommendations for operators and regulators

By **Nick Doty** & **Mallory Knodel**

We are grateful for insightful reviews from colleagues in civil society, academia, and the tech and telecommunications industries. Thanks to Stan Adams and Avery Gardiner for their research on network slicing and 5G during their time at CDT.

Footnotes in this report include original links as well as links archived and shortened by the Perma.cc service. The Perma.cc links also contain information on the date of retrieval and archive.

# Content

# Introduction

Since the 1990s, the Center for Democracy & Technology (CDT) has been concerned with the ways in which "service delivery models of the Internet risk diminishing or eliminating the rough 'equality of voice' between small and large speakers that is a key characteristic of the narrowband Internet."[1] The principles of net neutrality have been essential for maintaining the diversity of services built on top of the internet and for maintaining some competition between small and large providers of those online services. That diversity and competition, in turn, provide users with a broader array of choices for seeking online content and disseminating their own speech. Furthermore, in order for the internet to be used to its full potential and to protect the human rights of internet users, we need privacy from surveillance and unwarranted data collection by governments, network providers, and edge providers.

What has changed since the 1990s? In addition to the explosive growth of the internet, its worldwide accessibility, and its vital importance to conducting activities of modern life, new networking technologies provide capabilities for novel internet services as well as opportunities for network operators to monetize networks further.

One such change, the transition to 5G mobile networks, enables network operators to engage in a technique called network slicing. The portion of a network that is sliced can be used to provide a suite of different service offerings, each tailored to specific purposes,

---

[1] Morris Jr., J. B., & Berman, J. (2010). The Broadband Internet: The End of the Equal Voice? Center for Democracy & Technology. [perma.cc/44EF-BX33]

instead of a single, general-purpose subscription for mobile voice and data.[2] Moving from general-purpose access to a sliced network means treating network traffic differently, which, by definition, violates the strictest definition of net neutrality — that every data packet should be treated equally and identically. However net neutrality in practice does allow for differential treatment, "so long as it is not discriminatory in ways that affect the internet user's Quality of Experience or competition among edge providers."[3]

This requires a careful approach. Our report describes the technologies used for network slicing and outlines recommendations for an approach – for both operators and regulators – to enable network slicing while maintaining network neutrality, protecting privacy, and promoting competition.

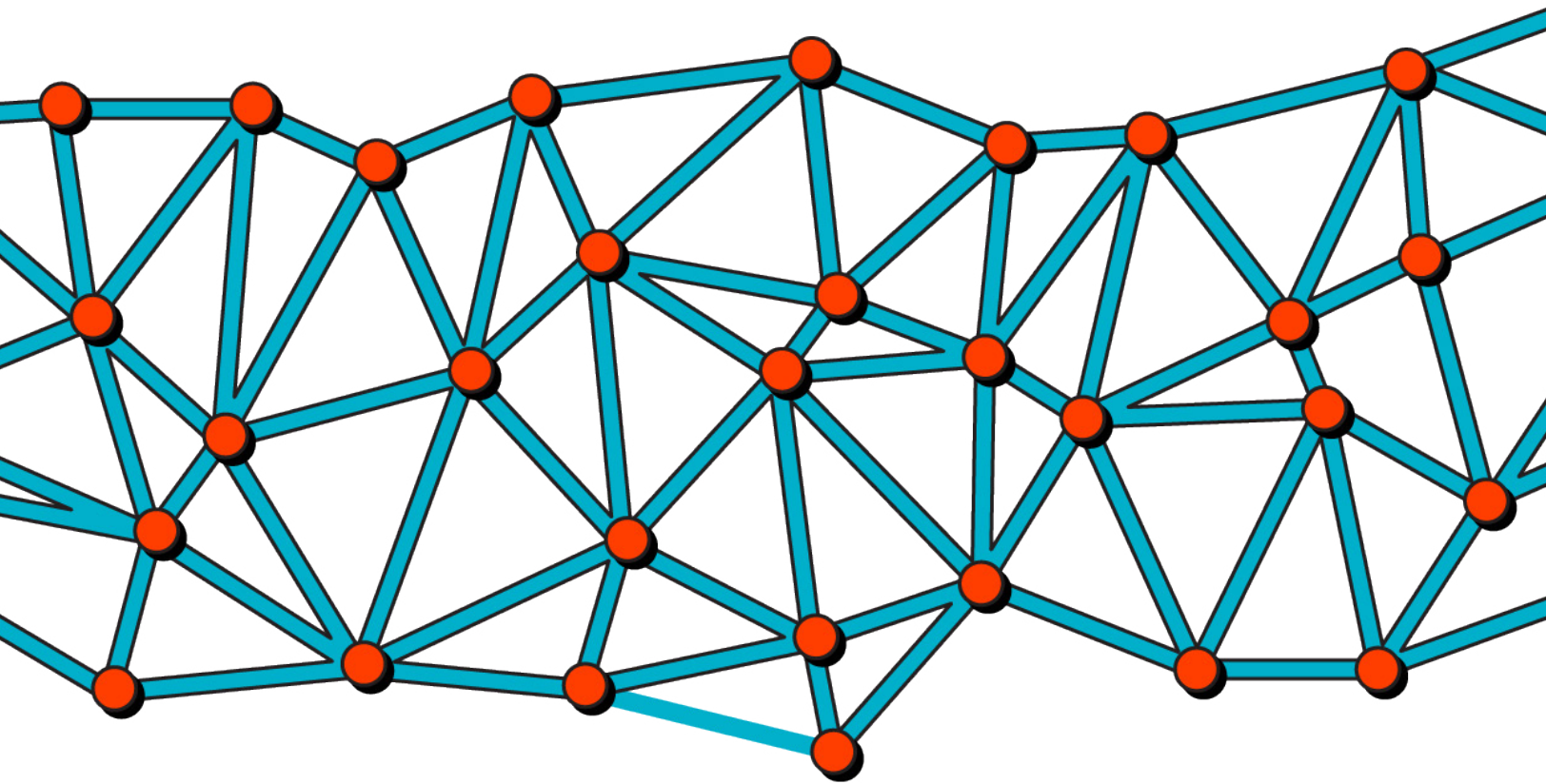

---

# Network Slicing

**N**etwork slicing creates customized, logically defined sub-networks to offer particular service characteristics between internet service providers and their customers' network-supported applications. Each "slice" consists of a virtualized network configured to meet requirements, such as access control, bandwidth, caching, edge computing, or latency. In effect, each slice is its own network, which might specify which endpoints, applications, or types of traffic the network allows. Telecommunications providers can sell these specialized networks to business with the promise that they are better able to support the performance of the business's desired applications and the preferences of their user bases. Network slices are better able to support this optimization – including support for new applications – than a "best-effort" network might be, where every type of traffic generally receives equal treatment.

Particular applications of this combination of techniques may vary and will depend on the particular technical details and business cases, as this report will describe further. But the prototypical examples include slicing a network (i.e. radio spectrum, bandwidth, and local computing resources) to include not only mobile broadband, but also certain high-reliability and low-latency applications – perhaps messaging between drones, self-driving cars, and city infrastructure in an urban environment – and also low-traffic applications that could accommodate high latency (such as reporting data from a high number of low-energy embedded sensors).

To fully support the specialized needs of applications, these virtual network slices must fully extend between the network endpoints using and providing the slices, which in many cases will involve several networks operated by different entities. Although the necessary standard-setting processes have been underway for years, it is not yet clear how competing network operators may choose to coordinate the performance of network slices to enable end-to-end functionality. As a result, initial rollouts of 5G network slicing may focus on use cases where the relevant endpoints are all on a single network.  Network slicing is likely to be used primarily to meet the needs of commercial, industry, and government customers. However, individual users will still be impacted by these service models, whether through direct subscriptions to internet service or through the effects on institutional customers like governments or health care providers. Over time, network slicing may become ubiquitous and central to the architecture of telecommunications standards such as 5G.

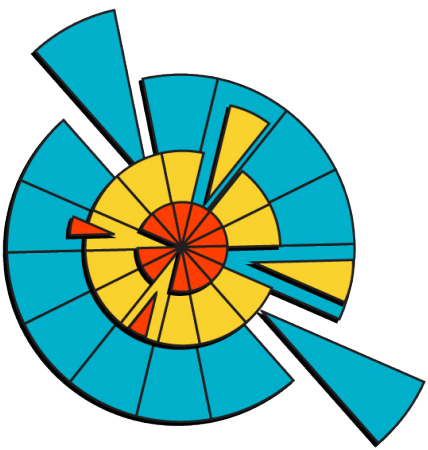

Aside from the technical logistics, the prospect of network operators coordinating on quality (such as service level, speed and latency) and pricing to provide end-to-end network slices may raise significant competition concerns, creating incentives to collude, exclude less powerful operators, and consolidate, all of which could limit choice and dampen innovation.

## Techniques of Network Slicing

Network slicing is the general approach of slicing up a network into customized networks with different properties for different applications.[4] Network providers will use different technologies to accomplish network slicing, which may include: **software-defined networking** to separate control and data transfer; **mobile edge computing** to add some direct computation functionality to the edge of networks; **virtualized network functions** to abstract and combine different networking components; and **orchestration** to dynamically combine and deploy these network configurations.

---

[4] For one succinct definition, see the 3GPP documentation, which standardizes 5G cellular communications:
> "A logical network that provides specific network capabilities and network characteristics"

3GPP. (2022). System Architecture for the 5G System (Technical Specification (TS) 23.501 V15.13.0). 3rd Generation Partnership Project. [perma.cc/ZK9K-QSZT]

## Software-Defined Networking

Historically, networking hardware has been proprietary and static—dedicated to a single function or limited set of functions. In engineer-speak, older networking hardware merged together the "control plane" (makes a decision about how to route information) and the "data plane" (sends information along the correct route).[5] This arrangement optimized for performance, but these pieces of equipment could only perform a fixed set of tasks. A network operator could not reprogram them to do something else or to do what they already do slightly differently. This meant that changing how a network worked was difficult, time-consuming, and expensive because the operator would have to replace or reconfigure each piece of equipment involved. Early telecommunications switches were called "large, immortal machines" because of their longevity, and that lack of flexibility may contribute to insecurity.[6] Middleboxes were a technical precursor to network slicing. They are networking devices that do more than simple packet forwarding, often used to manage network traffic or handle the decreasing supply of IP addresses. The mass deployment of this static network hardware contributed to the ossification of the internet, making it harder to adopt new protocols.[7]

With software-defined networking (SDN), the control plane is separated from the data plane. This means that a network operator can reprogram the control plane according to a variety of considerations and possible configurations, while leaving the data

---

[5] Wikipedia provides an accessible overview of the control plane and forwarding plane.

[6] Landau, S. (2013). The Large Immortal Machine and the Ticking Time Bomb. Journal on Telecommunications & High Technology Law, 11(1). [perma.cc/8LZK-5HWB]

[7] Middleboxes have been defined, categorized and debated at the IETF. Brim, S. W., & Carpenter, B. E. (2002). Middleboxes: Taxonomy and Issues (RFC 3234). Internet Engineering Task Force. [perma.cc/S54D-MSXP]

While it is hoped that SDN may improve the flexibility of middleboxes, the internet community also explores more explicit mechanisms for signaling to network intermediaries or encrypting communications from intermediaries.

Trammell, B., & Kühlewind, M. (2015). Report from the IAB Workshop on Stack Evolution in a Middlebox Internet (SEMI) (RFC 7663). Internet Engineering Task Force. [perma.cc/VYY7-PKL5] plane to simply receive and forward information. The control plane of each of the many, many pieces of networking equipment distributed throughout the network can be reprogrammed from a central location. This enables network operators to change the way their networks handle and route traffic more quickly and efficiently, which means that they can respond to changes in network usage as they happen, rather than changes taking days or weeks.

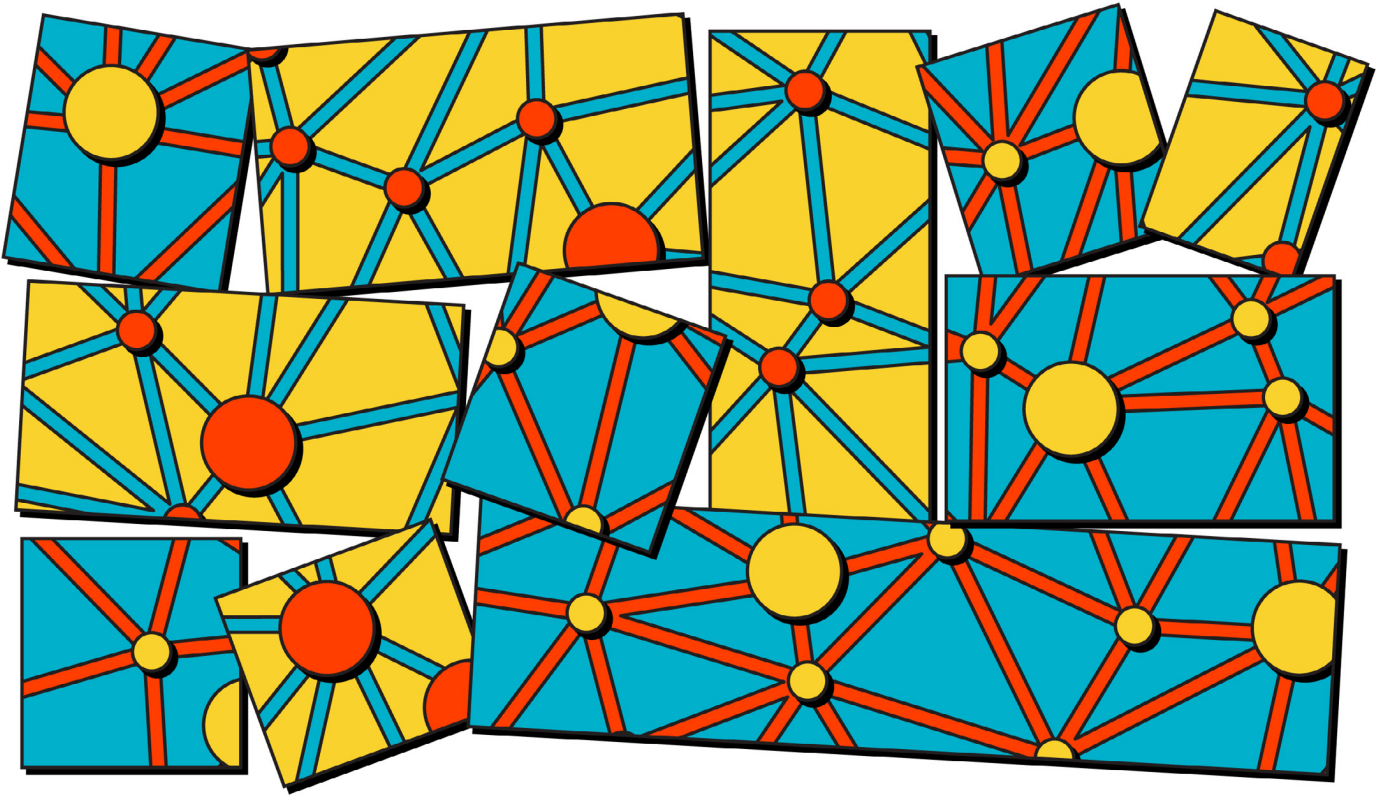

In some ways, the transition to SDN may be thought of as analogous to our transition from a bevy of dedicated purpose gadgets (camera, phone, GPS unit, calculator, etc) to a programmable device capable of performing many functions (a smartphone or tablet with different apps). The apps are like the new control plane instructions (you can create new or different ones depending on your needs), while the operating system and processor are like the data plane (they will carry out the app instructions). Sophisticated SDN even allows the rapid deployment of multiple software component packages, a little like the apps on your phone, so that operators can more easily "plug and play" their SDN modules.

On its own, SDN has real potential to make networks more agile and dynamic by making it cheaper, easier, and faster for network operators to reconfigure network node hardware. But when used to support computing and processing in the network, SDN turns network hardware into something more than communication conduits, and those increased capabilities could also potentially be used in ways that would violate network neutrality and have implications for privacy, too.

### Mobile Edge Computing

Mobile edge computing (MEC) is another emerging trend in communications networks in which small computers are distributed all around the "edge" of mobile networks. The "edge" means the part of the network sending traffic directly to users, like cell antennas or base stations—not between cell towers or between other elements of the "core" of the internet. These computing stations are much, much smaller (and have far less computing power) than modern cloud data centers, but they offer a different advantage: they are much closer to users or whatever device might connect to the mobile network. This means that information sent for processing that can be performed by the computing devices at the edge of the network doesn't need to travel as far and latency (overall trip time) can be lower, making the network very snappy. Those snappy, low-latency communications are a crucial part of some existing and emerging applications, such as augmented and virtual reality (AR, VR) and some kinds of autonomous vehicle controls, such as driverless vehicles communicating with city infrastructure like traffic lights.

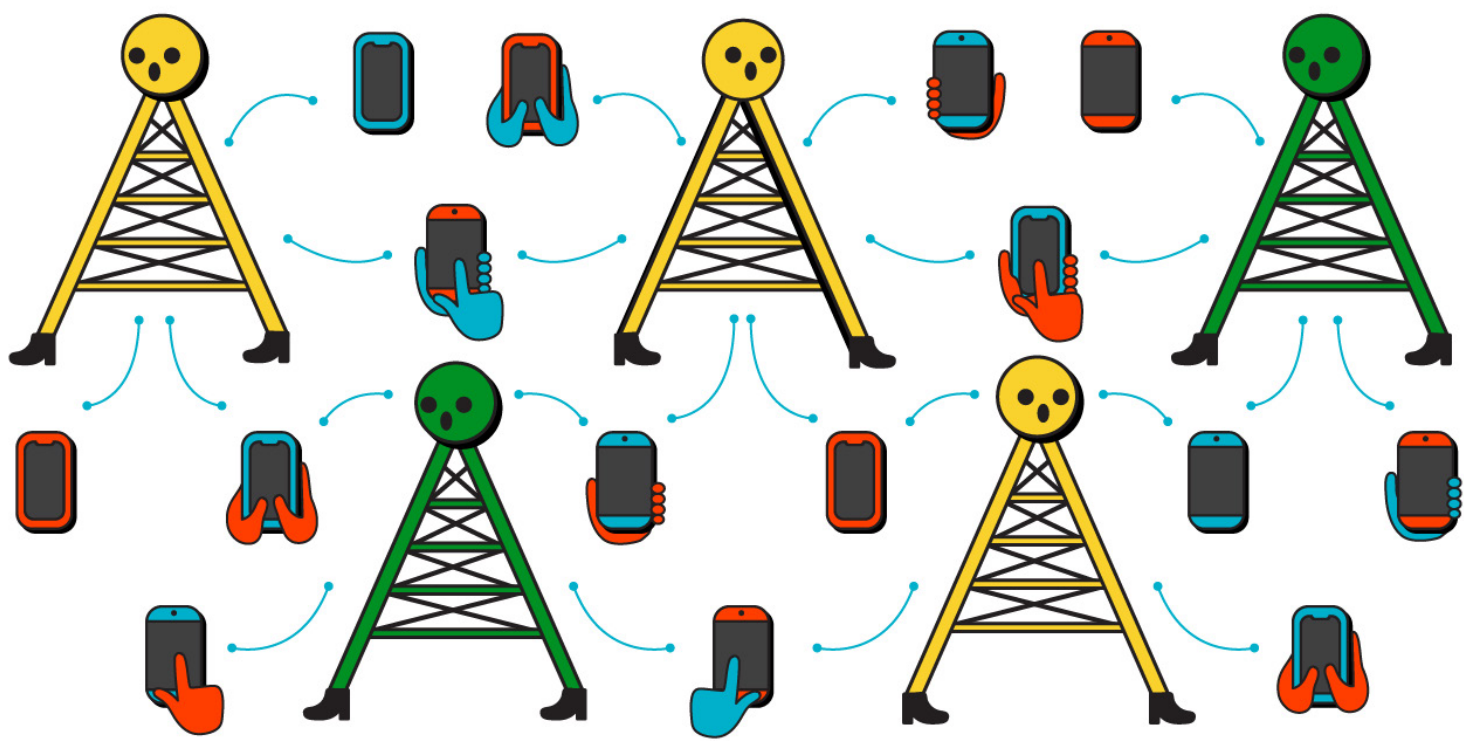

In addition to distributing service provision to the edges, MEC also means that the network can perform some kinds of computation instead of the mobile devices doing them, potentially reducing the need for powerful on-device computing and increasing battery life. It is difficult to predict all potential impacts of this capability, but it could lead to smaller, lighter, and cheaper devices. For example, a portable kiosk-style device would only need enough battery and computing power to send/receive wireless signals and display the results but could offer the processing power of one or more edge computing nodes (likely more powerful than current mobile devices). This could allow devices to be smaller, lighter, and more simply designed, while also employing the computing power of the network. However, MEC and other forms of computing in the network also raise concerns about privacy, security, and competition.

## Virtualized Network Functions

Virtualization, or creating a virtual "thing," involves modeling and running that thing in software instead of hardware: in bits instead of atoms. This software-based virtualized thing runs on top of the physical hardware in which it resides—for example, a virtual computer within a physical computer, or a virtual network within a physical network. This is how much of cloud computing works—a program sets aside a certain amount of processing power and other resources (usually on computers located somewhere else) and that chunk works as a standalone, virtual machine. There can be many virtual machines operating on a single hardware unit, or a single virtual machine utilizing the hardware resources of many physical machines. In some ways, virtual computing is like a fantasy sports team—those players don't physically play on the same team together (the pieces don't all reside in the same computer), but function as a team when assembled.

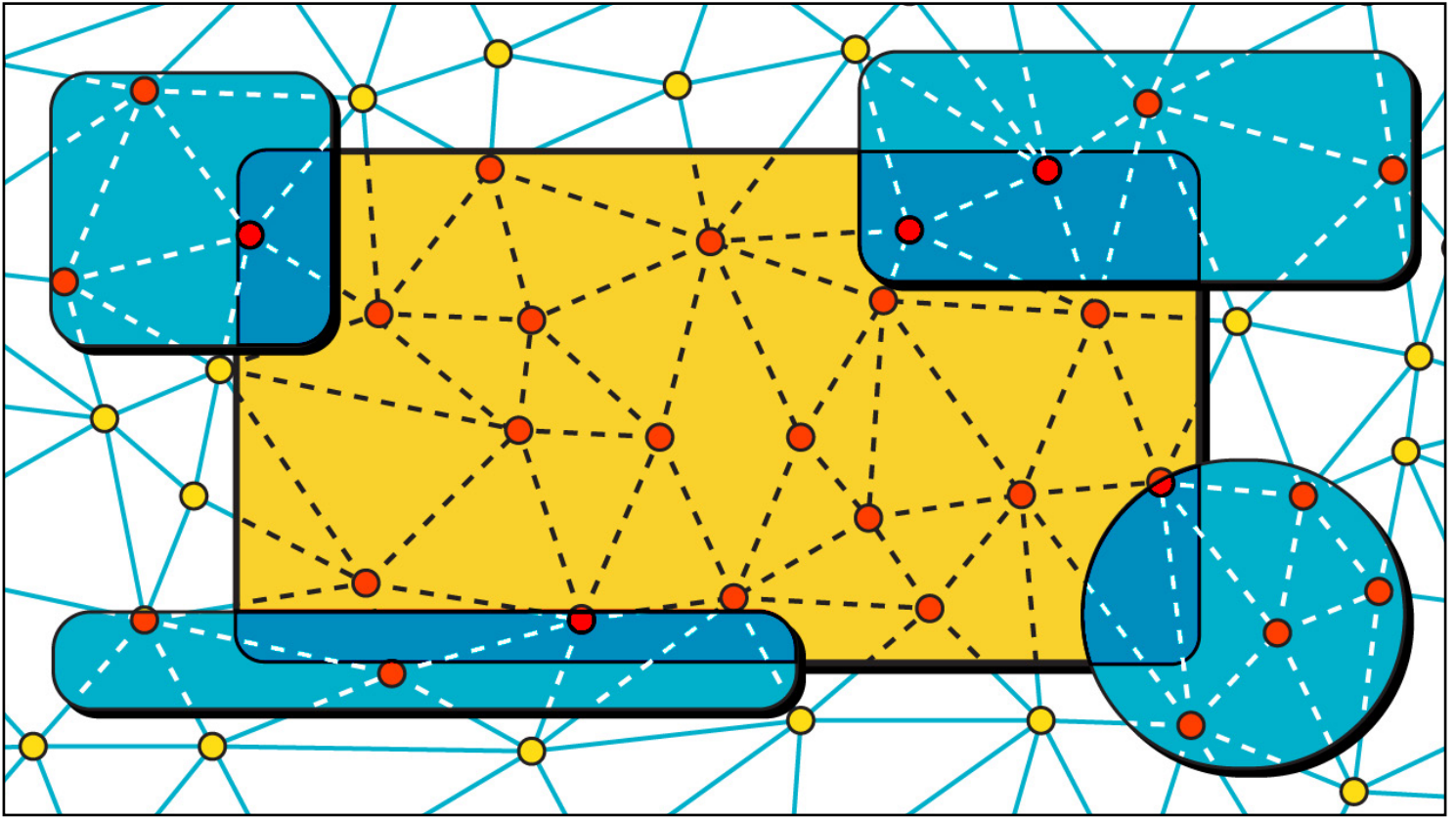

A virtual private network (VPN) is one common example of a virtualized network.[8] In that context, the "thing" being virtualized is a network, only it is not a physical network of wires, boxes, and radio transmitters, but rather a network that is logically separated—not physically separate wires, just separate connection rules and typically an encrypted tunnel—from other network traffic traveling on the same physical networks. Many enterprises use VPNs so that employees can connect to various services on the corporate network even when working from home while maintaining security and privacy by logically isolating those communications from other traffic at the employee's home or the ISP's network.

A virtualized network function, then, is a thing or set of things that a network can do, except instead of using a dedicated-purpose physical device, the function is enabled by software and uses a variable mix of network resources to achieve its purpose. It is possible to implement *network functions virtualization* (NFV) without using software-defined networking, but SDN gives operators the ability to adjust, move, remove, reconfigure, and change the scale of virtualized network functions from a centralized orchestration platform. So if a network operator wants to create a firewall between two segments of a network (to block harmful traffic from one side or the other), it can send instructions to the general-purpose unit at the desired node to describe what the firewall should do, how to operate it, and for how long.

These virtualized functions can be customized or even strung together to create a custom service that the network, or part of the network, provides. For traditional network functions such as moving information from place to place, controlling congestion, or creating secure paths, SDN and NFV help operators finely tune their networks and adjust them dynamically to meet different demands. But network operators can add other gear to their networks, too, such as data servers and general-purpose computers. In fact, many networks already have these resources built into them. For example, some access networks and core mobile networks incorporate racks of servers to support the local caching and delivery of popular content.

## Orchestration

Making customized mixtures of a network's virtualized components depends on the network operator's ability to orchestrate the assembly, use, and retirement of components in a modular fashion. Developers of other software-based systems have adopted various orchestration tools which allow them to construct virtual systems by compiling pre-assembled functional modules. These modules, called containers or microservices, each perform a discrete set of functions (like running some blocks of code) and can be linked together to perform more complex tasks. Because these containers are virtual rather than physical, they can be used, interchanged, and discarded quickly and easily, allowing operators to spin up and adjust complex virtual assemblies on short notice. Orchestration makes the configuration or reconfiguration of custom network slices simple and efficient, allowing operators to offer custom slices to meet needs as they arrive.

Each virtual network slice will be a mix of different network resources, including elements such as logically separated routing functions, in-network computing and storage, and portions of radio frequency channels. The exact "recipes" for individual slices will be assembled through some form of orchestration, as discussed previously, which will also determine when slices will be available, how long they will exist, and what types or sources of traffic they may carry.

As network operators move increasingly toward software-defined networking, their use of container orchestration will likely increase. However, because some of the benefits promised by network slicing may require continuous service levels across multiple networks – sometimes referred to as end-to-end network slicing – operators may also seek some way to coordinate their software-defined network slices across network boundaries. It is not yet clear whether and how operators will achieve this coordination, but a shared orchestration regime offers one possibility. As an alternative, operators may choose to create voluntary standards for popular slice configurations so that performance guarantees can be supported across networks. A third option may be that network operators formulate agreements on how they will share access to, and control over, physical or virtual components of their networks so that a single operator may be able to extend and control one or more network slices across multiple networks.

## Comparing Radio to Wired Networks

Network slicing offers potential benefits in the form of network efficiency. Rather than having all network traffic compete for the same general resources, slicing allows operators to better allocate limited resources so that different uses of the network only use the resources necessary to support their function. This is especially useful in radio networks because of limited spectrum availability – slicing helps operators get the most out of the spectrum they use by isolating different kinds of traffic and matching those streams with the best combination of frequency and channel width to support the needs of a particular use case.

For example, operators can set aside very narrow channels for certain uses, such as sensor arrays that transmit small amounts of information at regular intervals, which allows operators to utilize portions of the spectrum that might otherwise act as mere buffers between wider channels. Likewise, slicing might enable more nuanced control over how different kinds of network traffic are synchronized through time, space, and code division methods. Essentially, because slicing could be used to isolate different kinds of traffic, the traffic in any given slice would be more homogeneous, which could enable more efficient transmission across networks compared to the one-size-fits-most nature of best-effort network operation.

While spectrum usage makes network slicing especially appealing for wireless networks, it can also be applied to wired networks. A network slice could allocate certain bandwidth or network priority, and access to close-by computing resources, to a certain type or source of network traffic.

## Future Use Cases

How can network slicing – including software-defined networking, mobile edge computing, and virtualized network functions, deployed through orchestration – be used, and how do we expect it will be used?

Academic literature and marketing materials about 5G networks categorize three of the most prototypical categories of network usage: mobile broadband, reliable low-latency communications, and machine-type communication. **Mobile broadband** (or enhanced mobile broadband eMBB) envisions the uses we might be familiar with today in the use of our smartphones or laptops: browsing the Web, downloading media and applications, streaming video and communicating with chat, voice, and video, potentially at higher speeds than currently available in cellular connections. **Reliable low-latency communications** (or ultra-reliable low latency communications, uRLLC) would provide higher guarantees for reliability and latency, but potentially with lower bandwidth, and might be especially relevant for augmented reality, or control of drones, self-driving vehicles, or industrial robotics. **Machine-type communication** (or massive machine-type communication, mMTC) covers sensors and devices we often think of as the Internet of Things (IoT) or smart home or smart city devices. Machine-type communication use cases have needs for a very large number of devices in a small geographic area, with limited processing capabilities and low power usage, and often send messages that are not time-sensitive.

While these prototypical categories are illustrative, proposals for network slicing have extended far beyond these three generic categories.[9] Network slices could be made specific to an application or a particular service: for example, your phone could connect to multiple network slices, including a slice just for your bank, with the alleged security advantage that network traffic would go more directly to the bank's servers with fewer intermediaries. Network operators can provide slices with less congestion to customers willing to pay extra for a speed "boost" at a particular time, in a crowded location, or for a specific application. Slices could be customized to individual users or groups of users with similar network usage. Or slices could be provided for different classes of applications, with gaming or video conferencing having certain properties that could benefit from it. However, the adopted uses of network slicing remain uncertain at this more exploratory stage, and highly specific and end-to-end network slices appear less likely to be deployed, particularly in the short term.

Network operators hope to provide flexible, on-demand access to a suite of slices with different capabilities. They will likely seek to monetize these capabilities by offering them to providers of network-based services, business and other institutional customers, and potentially ISP subscribers, with the promise that the virtual network slice will outperform an ordinary internet connection with respect to speed, latency, reliability or some service element.

Virtualized networks that allocate different resources (spectrum, bandwidth, priority, edge computing) and are deployed and managed through dynamic orchestration split up the network, orthogonal to the layers of the internet. Network slicing can be used to divide the network into segments: either for different users or their different uses of data, for different endpoint services, or both. The implications for neutrality, privacy, competition, and innovation will depend on how this new class of technology is used.

---

[9] Some relevant marketing materials and other literature suggest possible use cases that technology or service providers intend for network slicing use, though that doesn't guarantee their actual adoption or widespread use.
- Taleb, T., Mada, B., Corici, M.-I., Nakao, A., & Flinck, H. (2017). PERMIT: Network slicing for personalized 5G mobile telecommunications. IEEE Communications Magazine, 55(5), 88–93. [perma.cc/J8ZQ-PAC3]
- Ericsson. Dynamic End-user Boost. [perma.cc/K7ZF-YN2Q]
- Ericsson. (2021, November 1). FarEasTone and Ericsson mark a breakthrough in 5G network slicing. [perma.cc/GEA4-DQ2S]
- GSMA. (2020). An Introduction to Network Slicing. [perma.cc/6BAS-PWDK]

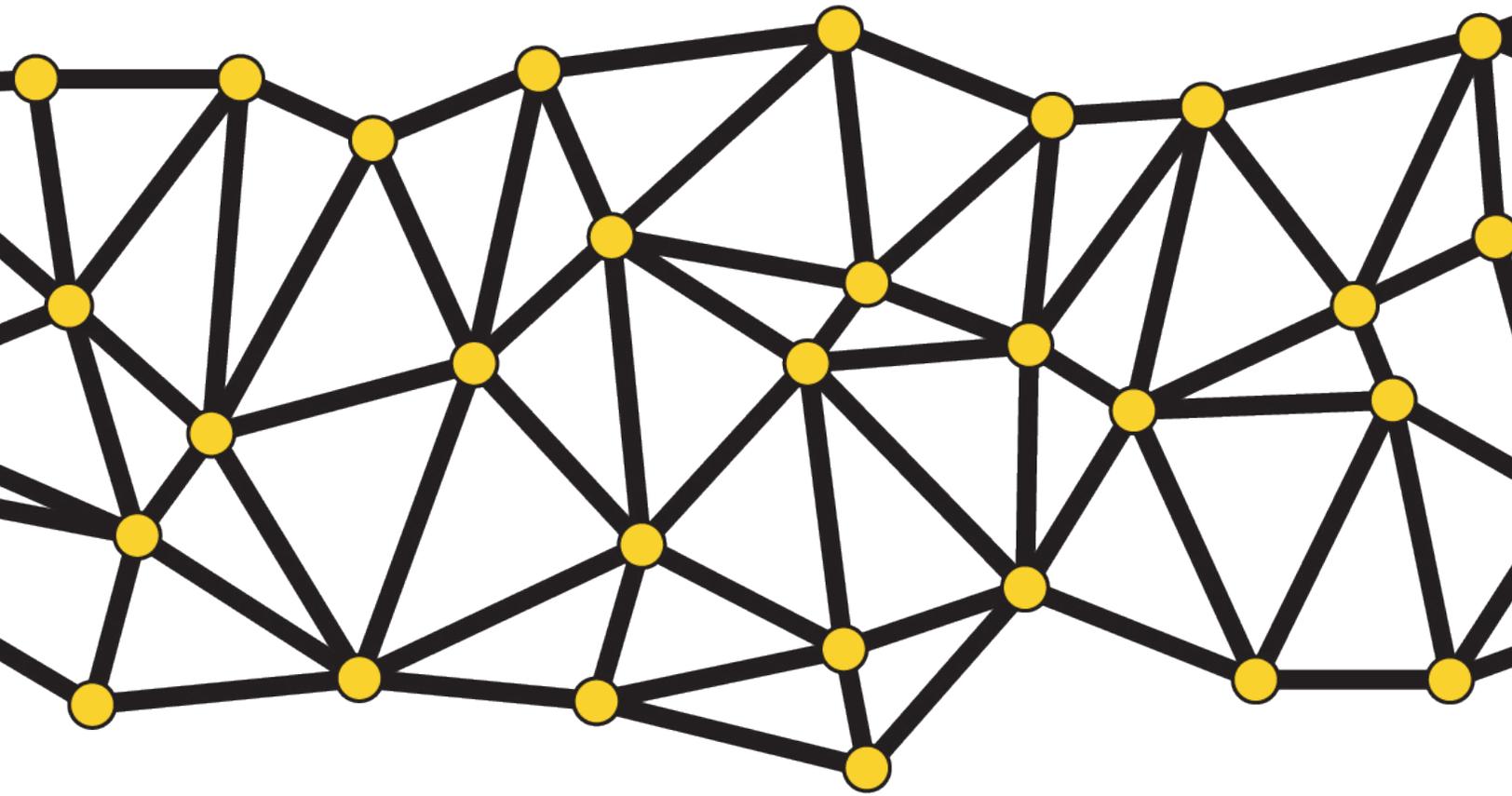

# Network Slicing and the Open Internet

### Net Neutrality

Network slicing is an exciting concept, but its capabilities also raise several concerns about its potential impact on the ideals of the open internet. First, creating a practical way to consistently treat some kinds of traffic differently than others, and to preserve that distinction across multiple networks, means that network operators will be more capable of favoring or disfavoring certain traffic based on its source, destination, protocol, originating device or application. Second, the ability to charge premiums for virtual network slices capable of providing higher qualities of service or experience potentially creates perverse incentives for network operators: the worse their "standard" offering performs, the more appealing the premium slices appear. Third, depending on the application of and demand for slices, operators may wish to devote increasingly large portions of their overall network resources to supporting premium slices, which could cannibalize the capacity available for general-purpose internet access.

Without appropriate safeguards in place, the combined effects of these possible outcomes may reduce both the openness of the internet and the choices of content, applications, and services available to internet users.

### Discriminatory treatment

Although the act of subdividing a network into multiple virtual networks does not, on its own, imply that the data passing through them will be subject to discriminatory treatment policies, the ability to treat data differently is the primary purpose of network slicing. Through slicing, networks can be configured so that they are best suited to support the service-level needs of different kinds of data traffic. Individual slices would likely be both exclusionary (only certain kinds or sources of traffic will be allowed) and discriminatory (the network's configuration will have different impacts on the performance of the applications whose data the network transmits).

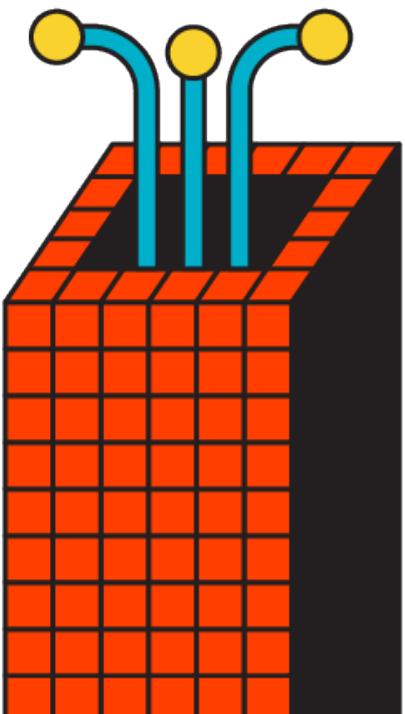

When applied through regulations, the principle of net neutrality has been subject to limitations and exceptions, such as allowances for certain specialized services and reasonable network management practices. Specialized services or "non-BIAS data services" are offerings by broadband providers that do not reach large parts of the internet, but are instead a specific application level service.[10] Network slicing may be just another way to implement a network operator's specialized services or network management policies. Operators could use slicing to offer specialized services or carry out reasonable management practices that do not create anticompetitive, discriminatory, or market-distorting outcomes. As we have argued previously, reasonable network management practices should typically be agnostic to the content or application.[11] However, operators could also use slicing to create new ways to favor their affiliates, to create artificial scarcity and higher prices, to create barriers to entry for new competitors, or other undesirable impacts on internet users. This does not mean that regulators should prohibit operators from using network slicing as a specialized service or a method to make networks more efficient, but the nature of the technology requires close oversight if open internet principles are to be preserved.

## Incentives on quality

Network slicing offers the ability to create specialized networks capable of outperforming best-effort networks by offering guaranteed levels of certain service characteristics. Although some applications simply cannot function without such specialized treatment, consistent service levels in mass-market, retail broadband can also improve the users' quality of experience for many other applications. For example, many popular video conference applications are functional today, but the user experience they offer could be improved if networks could guarantee consistently low levels of latency and higher levels of reliability, resulting in less laggy conversations and fewer frozen video feeds.

---

[10] See FCC Open internet Order, "Non-BIAS Data Services", paragraph 207. Federal Communications Commission. (2015). In the matter of protecting and promoting the open internet: Report and order on remand, declaratory ruling, and order (FCC-15-24). [perma.cc/WV7D-JJR2]

[11] CDT Comments to FCC on the Open Internet. (2014). Center for Democracy & Technology. [perma.cc/C5KF-BVAU]

To enhance the appeal and market value of premium network slices, operators could be motivated to demonstrate a strong contrast between the performance of network slices and best-effort network management. This could create perverse incentives to degrade the performance of best-effort networks, either actively or by neglect so that providers and users of Web applications will be motivated to opt for the more expensive network slice rather than suffer increasingly poor quality of experience on the best-effort network. And even if the performance of best-effort networks remains constant, the availability of network slices may create incentives among providers of some applications to opt for a higher-priced and better-performing network slice, if doing so distinguishes the performance of their applications from those of competitors. Because some attributes of network performance are to an extent rivalrous goods – exclusive properties that can't be shared by all users of the network – not every provider will be able to enjoy the same level of performance. This may mean that the largest, most established providers would be able to further entrench their market position if they were willing to outspend would-be competitors on network priority to ensure that their applications consistently outperform all others, harming the equivalency of voice of large and small providers.

## Cannibalization of Capacity

All broadband networks have limited capacity to support communications. Moreover, particular attributes of network performance, such as latency, can be negatively impacted as traffic volume increases. But network slicing offers a way to guarantee certain aspects of network performance without being impacted by other uses of the same physical network. That does not mean, however, that other uses of the physical network are not impacted by slicing. Where slicing divides up resources, including bandwidth or network capacity, and dedicates them to particular specialized networks, network slicing may cause decreases in the capacity available for best-effort networks. However, slicing may also be implemented in such a way that it doesn't decrease best-effort capacity, but instead uses resources (like spectrum or edge computing) more efficiently.

Although it is difficult to predict whether network slices will be used to support novel uses of communications networks or enhance the performance of existing uses, or both, the volume of traffic that networks transmit will surely increase. Because network operators will likely monetize network slicing by charging higher rates to providers or users who make use of premium slices, they will be motivated to use as much of their network capacity as required to meet the demand for such slices. Although most of the existing uses of the internet do not necessarily require the guaranteed levels of service that slices are designed to provide, deep-pocketed application providers may opt to pay for access to a network slice to provide a better quality of experience for their users. Combined, these incentives could lead to a greater portion of network capacity devoted to custom slices, leaving less for general-purpose uses of networks. This risk of slicing raises the long-held concern of segmenting some network traffic onto "a winding dirt road".[12]

## Competition

We expect competitors to compete, and network operators are no exception. In places where network operators face competing providers, they advertise heavily to individual and enterprise customers, touting quality and offering price promotions.

To make network slicing work across networks, competing providers must cooperate. If one network's slice is incompatible with another network's analogous slice, then the gains that network slicing could produce will be limited. Cooperation is necessary for the interoperability across networks that is characteristic of the internet. In addition to the coordination challenges, network slicing raises competition concerns, such as vertically-integrated providers facing incentives to prioritize their own services. Priority to particular uses of the network or particular services that have pre-arranged agreements with network providers may discourage generativity and choices of the kind we have seen, and hope to continue to be able to expect, from the internet.

---

[12] Lessig, L., & McChesney, R. W. (2006, June 8). No Tolls on The Internet. Washington Post. [perma.cc/X73H-26C5]

## Coordination and Exclusion

The antitrust law issues around coordination for network slicing are akin to those that would arise in any collaboration among competitors.[13] If the wireless and fixed broadband companies in any given market get together in a room to discuss how to make their network slices interoperate, there would be an opportunity for them to discuss other kinds of coordination that are forbidden under the antitrust laws, like where to offer service and how to price it. They might also agree on limitations on network slicing that are convenient for them but disadvantageous to consumers and the public. Similar issues could arise even when network providers who currently operate in different geographies (such as the United States and Asia) collaborate, as they could potentially enter each other's markets and offer competing alternatives.

There are also concerns about which competitors would be able to participate in standard-setting processes or other collaborative arrangements of cross-domain network slices. Large providers could, for example, exclude smaller providers from the conversations about interoperability. They might select protocols that disadvantage smaller competitors or even make it impossible for them to offer network slices, all in an effort to drive business to themselves. In the past, some efforts to standardize emerging technologies have raised these kinds of antitrust concerns.[14]

Effective and responsible standard-setting bodies have developed clear principles and concrete policies and processes to address these categories of concern, so that collaboration can promote competition rather than foreclose it.[15] Open processes are important to ensure the inclusion of the perspectives of different potential competitors, stakeholders in different industry sectors, and the voices of civil society and end users. Decision-making by broad consensus, with clear procedures and documentation of decisions, prevents the capture or distortion (or appearance of distortion) of standard-setting processes. Antitrust policies and guidance should include due process around decisions and appeals and evaluation of proposals, based on technical merit and the public interest. Finalized standards and regular drafts should be made publicly available for wide review, and policies should require fair, reasonable, non-discriminatory, and preferably royalty-free licensing so that an adopted standard does not become a source of anti-competitive required paid licensing of essential intellectual property. Finally, voluntary adoption of standards promotes marketplace competition and technological development.

---

[13] See, for example, these guidelines from the federal antitrust agencies. Federal Trade Commission & Department of Justice. (2000). Antitrust Guidelines for Collaborations Among Competitors. [perma.cc/F4UR-92PM]

[14] See the prominent examples of *FTC v. Qualcomm* and *FTC v. Rambus*.

## Vertical Integration

Network slicing will treat some packets of data differently than others. That is a fundamental benefit of network slicing: the ability to optimize for a specific application, such as enabling robotic remote surgery (which requires very low latency and a very stable network) or remote sensors for refrigerated products in trucks (which can tolerate lower service levels).

But treating data packets differently can also present opportunities for anticompetitive market distortions, especially with vertically integrated providers. For example, if a network provider also owns a streaming video service, it might try to create network slices that optimize for its content while treating a competitor's content unfavorably. Any network operator that has a separate business that uses the network could have similar incentives for self-preferencing. Indeed, a network need not actually own a separate business that uses the network for there to be opportunities for distortions based on self-interest; a contractual relationship could provide the same incentives. These arrangements may apply not just to the bandwidth, but also to the computing resources that are envisioned for network

---

[15] The OpenStand Paradigm is one example of standard-setting bodies describing these principles as a group, but similar practices or subsets of this list have been adopted in different standard-setting venues. [perma.cc/6FZH-ZPJ6]

slicing. Access to geographically-nearby edge computing resources could be offered openly to some, and in more limited ways to others, so as to provide a preference to a vertically-integrated operator.

Lastly, network slicing could lead to consolidation, either horizontally or vertically. From a horizontal perspective, owning more networks reduces the need to interoperate with a multitude of other networks, and deal-seekers could try to claim those as efficiencies. This would present competitive concerns in an environment that already has very few choices of network operators. From a vertical perspective, firms that want to offer services that take advantage of network slicing may see value in buying networks. This could result in the largest tech firms becoming even more dominant if they buy networks. This could warrant antitrust enforcement, or in its absence, regulatory action may be required.

## Impacts on User Choices

The incentives described above – maximizing slice utilization and increasing the value of slices relative to best-effort service – may lead popular online application providers to be willing to pay extra for network slices, and may lead network operators to charge more. Over time this could magnify the differences in the performance and popularity of those applications whose providers can afford and are willing to pay the higher rates charged for custom slices where network operators include dedicated slices in broadband internet service offerings. This may lead to the rise of a "premium" internet with limited offerings and a "standard" internet with reduced capacity to support all remaining internet uses. Whether ISPs charge application providers or users or both for the privilege of using network slices, the end result would be higher prices for internet users. Likewise, because the success of many kinds of online applications would depend on their providers' ability to pay for premium service, and because that premium service would be a limited resource, the choices available to internet users may dwindle as a few dominant applications prevail while smaller competitors languish in the "standard" internet. This could also hinder the emergence of innovative new applications.

It is possible, though perhaps unlikely, that ISPs would offer access to virtual network slices directly to consumers. Although there are circumstances in which such offerings would not create tension with open internet principles, they may follow models like those associated with differentiated pricing, or zero rating. Under these models, ISPs might establish arrangements with application providers to utilize network slices designed to enhance the performance of certain applications, and then advertise the enhanced performance as a benefit of using their network or a reason to pay for upgraded service. Either of these models would tend to reduce users' choices, for the reasons discussed previously. They would also create additional tensions because of the potential for exclusive arrangements and other means of market distortion because ISPs would be in a position to choose which applications or providers will be able to enjoy the enhanced performance of network slices. As we have previously noted in an analysis of zero rating configurations, exclusive, sponsored, discriminatory, or opaque arrangements can distort markets and limit users' choices of online services.[16]

User-facing network slicing access could also increase the opportunities for price discrimination: charging customers an extra fee to "boost" their network speed in a particular area or highly congested time, for example. Transparency may be especially important in those cases; more dynamic pricing may lead to increased costs or decreased quality for customers, not necessarily driven just by congestion but by unfair practices to extract additional revenue.

If they favor their own affiliates or simply cater to the highest bidders, ISPs could alter the competitive landscape for providers of online applications and services. The effects of this control could extend directly to internet users, whose choices would be constrained by the performance, price, and availability of applications available to them.

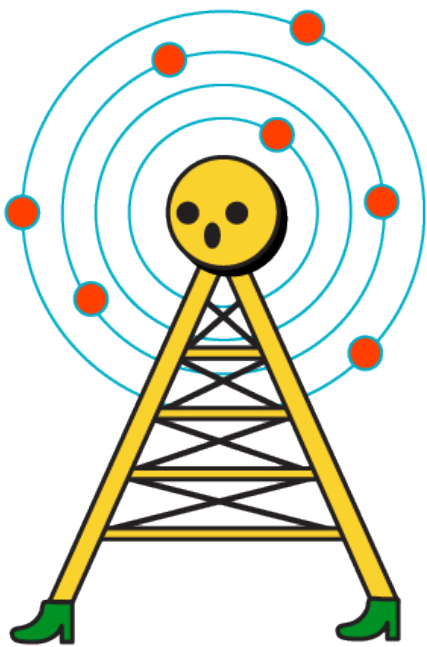

# Privacy and Access to Information

Privacy is essential for open communication: without secure and private methods, people will be chilled in how they use the internet or online services. At the network level, surveillance of communications also contributes to other privacy harms, including: profiling, targeting, and manipulation based on communications behavior; inferring social connections; controlling access to journalism or to dissenting political speech; and even identifying anonymous speakers and locating them. Because network access is so fundamental to the use of the internet – a necessary prerequisite, and at an underlying layer on top of which so many different services and applications are built – we should pay particular attention to the potential privacy impacts of new networking technology.

## Network Visibility and Surveillance

How network slicing and network virtualization in general will affect privacy online will depend on particular implementations. In some cases, network virtualization that is initiated and controlled by the user can provide security and privacy advantages: for example, VPNs can protect user privacy from network providers and other users on the same network, and give users a choice of whom to trust with analysis of their network traffic. But network slicing could also depend on the network provider having more detailed visibility into network usage. Rather than just delivering packets to an intended destination, the provider must determine which traffic should be assigned to which network slice, in order to assign that traffic priority regarding latency, spectrum, or bandwidth, or to connect to edge computing resources. What type of activity we engage in when we communicate online can reveal more than the general use of the internet: not just that you're sending or receiving network traffic to and from another computer, but whether you're conducting online gaming, video chat, telehealth, email, or chat, and what kind of devices you have connected (drones, vehicles, home automation sensors, etc.).

In the past, deep packet inspection has been proposed for various purposes by network providers, including for some kinds of efficiency in network management. As CDT has noted in the past, deep packet inspection provides invasive and dangerous access to the contents and details of communications and, even when intended purely for beneficial purposes, fundamentally threatens the privacy of internet users.[17]

The exact privacy implications of different network slicing configurations will depend on their implementation. But notably, even network slicing without highly-customized personalization, and with controls on the user device, may be revealing. For example, if a network provider provides a profile for download to the user device with some suggested configurations for video conferencing, gaming, and email, the use of that configuration will provide additional information to the network provider on that user's usage habits, even if the network provider does not engage in content-level analysis in order to select network slices.

## User Data and Meaningful Consent

Where the use of network slicing provides additional data on user activity, that data on how the internet is used may be sensitive. Internet users are often not direct customers of the network provider, even if their family, company, or friends have a direct account with the ISP. And for many users of the internet, the network provider is unknown or invisible, compared to the awareness one might have about which piece of software one is using or which website one is visiting. Network slicing that is coordinated across networks may also provide some of that insight to other network providers where the end-user has no customer relationship at all. As internet access is now a functional necessity – for work, school, political and social life – privacy cannot be meaningfully protected just by giving users the alleged choice to forgo the use of the internet altogether.

While baseline privacy legislation is necessary across jurisdictions as a

---

[17] Chief Computer Scientist Alissa Cooper made this case in 2008.
What Your Broadband Provider Knows About Your Web Use: Deep Packet Inspection and Communications Laws and Policies, House Committee on Energy and Commerce, Subcommittee on Telecommunications and the internet, 110th Congress (2008). [perma.cc/3QQE-38VS]
But see also: Communications Networks and Consumer Privacy: Recent Developments, House Committee on Energy and Commerce, Subcommittee on Communications, Technology, and the internet, 111th Congress (2009) (testimony of Leslie Harris). [perma.cc/8MAH-FESL]
Llansó, E., & Cooper, A. (2012, November 28). Adoption of Traffic Sniffing Standard Fans WCIT Flames. Center for Democracy & Technology. [perma.cc/Q8J6-3P24] basic protection, there may be special needs for telecommunications regulators to work with consumer protection regulators to provide meaningful privacy protections when new networking technologies may lead to additional visibility or data collection, especially when meaningful consent is infeasible.

## Architectural Designs for Protecting Privacy and Circumventing Censorship

In considering privacy implications, we should also look at trends in internet architecture. The end-to-end principle – that functional applications of networks are best implemented as much as possible by the endpoints rather than by the intermediary network – has historically contributed to the diverse, generative success of the internet as a platform.[18] Internet routing presents some significant challenges for privacy and censorship resistance because of the necessity of communicating to intermediaries what endpoint the traffic is intended to reach. But internet engineers also have the capability to take advantage of that end-to-end principle to improve upon privacy and censorship resistance. As we consider new network technologies, we must address where they could interfere with those long-term trends.

Proxies provide for protecting the identifying source address (an IP address, say) from the target service (the website you're visiting); they can also hide the destination from the local network provider. This category of technology includes "onion routing," researched in the 1990s and made available in the Tor network. But Apple has also provided its iCloud Private Relay service with some similar privacy properties based on MASQUE,[19] and we see growing interest in

---

[18] The end-to-end principle has a long history (and unclear definition) at the Internet Engineering Task Force.
Kempf, J., IAB, & Austein, R. (2004). The Rise of the Middle and the Future of End-to-End: Reflections on the Evolution of the internet Architecture (RFC 3724). Internet Engineering Task Force. [perma.cc/8DGC-4HCJ]
For one summary of how the end-to-end principle protects users from harm, see RFC 8890.
Nottingham, M. (2020). The Internet is for End Users (RFC 8890). Internet Engineering Task Force. [perma.cc/F75B-GBVH]

[19] See RFC 9298 and other documents from the MASQUE Working Group on proxying Web traffic over encrypted channels. Schinazi, D. (2022). Proxying UDP in HTTP (RFC 9298). Internet Engineering Task Force. [perma.cc/WG7X-NN8X]

privacy-preserving proxy technology.

These classes of technology are also useful for censorship resistance. The ability to hide the ultimate destination and contents of your communications from your network provider is necessary for those under authoritarian regimes to evade the filters used to censor unwanted political speech or external information.

End-to-end network slicing may come in conflict with the end-to-end principle of internet architecture and these technologies for privacy and censorship circumvention. Network slicing relies on some visibility of the application or destination endpoint to the network provider in order to use the techniques of network slicing described previously, selecting traffic for the right slice so that it receives the corresponding computing and network resources. Proxies that protect privacy or circumvent censorship specifically hide the end application (and as much other information as possible) from the network provider by routing traffic through a proxy or series of proxies that are not specific to a particular endpoint or application. Those mechanisms can continue to function through a general internet slice, but would not be able to take advantage of network slicing that is specific to particular endpoints or uses. More extensive use of network slicing could inhibit further adoption of privacy and internet freedom technology.

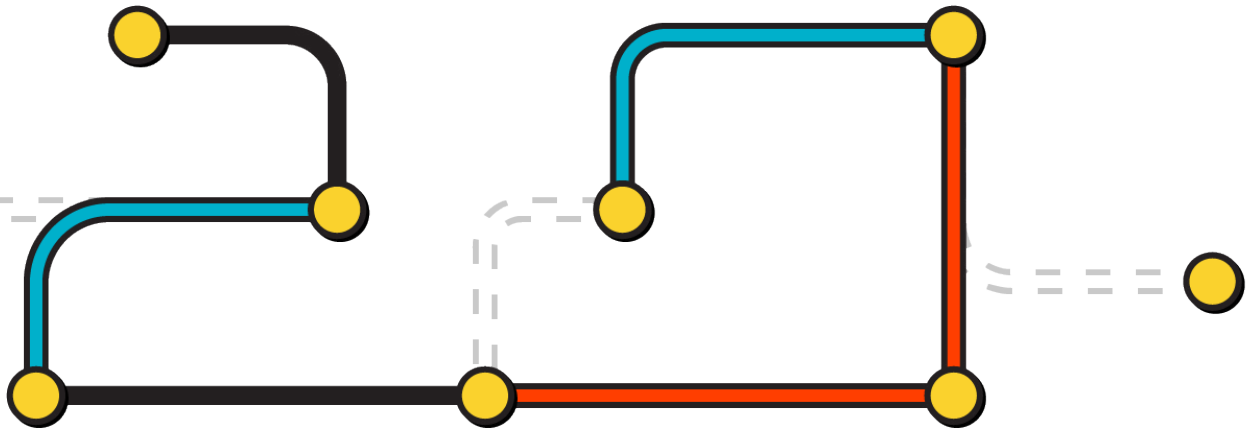

# Regulatory Approaches to Network Neutrality

Open internet principles like network neutrality enjoy regulatory protection in many jurisdictions, globally. Regulatory approaches vary, but some existing approaches may be better suited than others to address concerns raised by network slicing. In some cases, network operators have voiced concerns that regulations protecting the open internet create uncertainty or burdens for their anticipated deployments of network slicing systems.[20] But some regulators have noted how existing principles of net neutrality regulations can still be applied to network slicing and future developments.[21]

Network slicing may be used for certain models of monetization that conflict with open internet principles and the technique can also have implications for privacy and competition. But the technology itself can also be used in beneficial ways.

We urge regulators to review their approaches to preserving net neutrality in light of new capabilities such as slicing. If those approaches are insufficient to address the concerns raised by slicing or other emerging technologies, regulators should strive to address those concerns in a technologically agnostic manner and avoid overly narrow or prescriptive policies. We discuss some of the existing regulatory approaches that might be adapted to address network slicing.

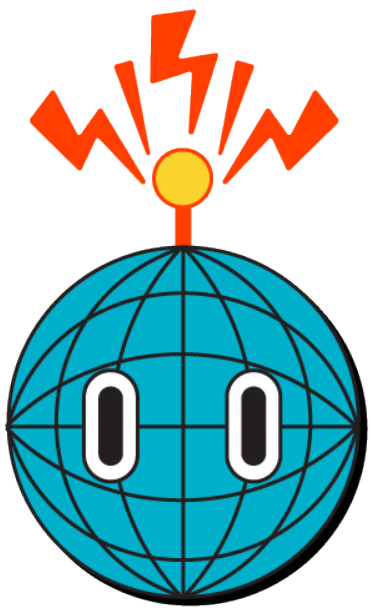

## Specialized Services vs. Paid Prioritization

As net neutrality principles have been refined through time and practice, it has become clear that an absolute view of net neutrality – where all internet traffic is treated identically – is impractical. Instead, network operators are encouraged to practice traffic management techniques that improve the quality of experience for classes of applications, so long as such techniques do not degrade the quality of experience for any users, and even continue to improve it. Under such management techniques, operators may prioritize the transmission of certain kinds of traffic, such as real-time video feeds, through congested nodes to reduce the latency of those communications. In many cases, it is possible to do this without detracting from the quality of experience of any network user, because other kinds of applications, like email or Web browsing, are less sensitive to increased latency. However, if ISPs are allowed to profit from giving some traffic preferential treatment, then open internet principles are at risk.

Regulators in both the European Union and the United States have tried to strike this balance by addressing aspects of existing network management practices that are similar to the issues raised by network slicing. In particular, the concepts of "specialized services" and "paid prioritization" are relevant to a regulatory approach aligning the practice of network slicing with open internet principles.

The term "specialized services" generally refers to network-based services that require certain qualities of service that are not guaranteed by traditional best-effort service and that do not provide general internet access.[22] This term has been used to classify services that fall outside of net neutrality regulations so that they may be treated differently than other services without running afoul of the regulations. Applications like remote surgery and internet-based television broadcasts (IPTV) have been cited as examples of specialized services.

In some cases, regulators have defined such services as having "objectively different" requirements for qualities of service.[23] As a means of classifying a practice like slicing, this definitional approach can help regulators who wish to limit ISPs' ability to evade regulation by partitioning general internet access into a suite of limited-purpose services. The requirement that specialized services demonstrate service level needs "objectively different" from what best-effort networks can provide is crucial. If certain services are to enjoy different, preferential treatment by network operators, then providers must show a legitimate technical need and restrict the service that gets that different treatment to that purpose.

In its 2015 Open Internet Order, the U.S. Federal Communications Commission prohibited ISPs from engaging in "paid prioritization." Essentially, ISPs were forbidden from providing more favorable treatment to some traffic (such as from a particular application provider) for money or other forms of value. Although this prohibition did not apply to "non-BIAS" services (the equivalent of "specialized services"), it could be a useful approach for preventing some of the ways network slicing might impact open internet principles, in addition to a definitional approach using "specialized services."

Historically, paid prioritization has been more difficult to implement in practice than in theory because network operators have not been able to establish agreements about how they will treat "priority" traffic from other networks. That is, there is no guarantee that operators would respect the wishes of other operators in terms of how they treat prioritized traffic streams, and the benefits of prioritization would be partially or totally lost unless competing networks agree on prioritized treatment. But if operators can agree on a system to coordinate end-to-end network slices, they could be motivated to implement paid prioritization models to further monetize their networks.

With appropriate restrictions, the impacts of paid prioritization on open internet principles can be mitigated. For example, prohibiting the paid prioritization of traffic transmitted as part of general internet access service helps to preserve the flatness of the internet by ensuring that providers of Web applications cannot buy an advantage over their competitors.

Although paid prioritization may produce the same impact for "specialized services" offered separately from general internet access service, it may be desirable to allow paid prioritization in this limited context. Indeed, some functionality may depend on certain service level guarantees in order to function; offering these services based on best-effort would not be possible. Enabling the development of these services may produce benefits that outweigh the concerns raised by paid prioritization, especially if paid prioritization is only allowed for services other than internet access. However, creating a legal distinction between general internet access and specialized services requires regulators to carefully assess the technical requirements of any services purported to require special treatment. Otherwise, providers may make false claims about service level needs to evade regulation. Even specialized services with particular performance needs should be offered in non-discriminatory ways, so that network access to that class of services can be used fairly by competitors.

As a safeguard against the "cannibalization" of network capacity by specialized services or network slices, regulators should require network operators to maintain or grow the portion of their networks' capacity used for general-purpose internet access, commensurate with user demand. ISPs should continue to enhance the performance of their general-purpose access services rather than neglecting them to increase the appeal of network slices. Ongoing testing and data-sharing can provide a way to evaluate ongoing additional services. Opponents of net neutrality regulations have argued that market competition has and will continue to provide incentives to continually increase general internet service capacity;[24] monitoring and evaluation may help determine whether that assertion is well-founded in different markets and with different network technologies.

## Enterprise Services

We expect that the first commercial deployments of network slices will be in the context of enterprise-level or business service accounts. From an open internet perspective, deployment in this context may raise fewer concerns than deployments in consumer environments. However, regulators should still consider the impacts on competition and the potential for market distortion.

Where regulators have addressed similar concerns, such as those raised by the practice of "paid prioritization," they often do so through narrow exceptions to a broad rule ensuring neutrality toward internet traffic. For example, regulators in the EU, the U.S., and India carved out exceptions for services providing single-purpose or limited functions that also require certain qualities of service (beyond what best-effort networks offer) to function. For regulators with similarly structured net neutrality regulations, this may be a good option for addressing slicing, in a way that will apply to both enterprise and retail consumer services.

## Network Management

As regulators across the globe have begun to implement rules to preserve net neutrality, they have faced the challenge of structuring those rules so that network operators are permitted to undertake beneficial, and often necessary, traffic management methods to address problems such as temporary congestion and network-based

---

[24] Sidak, J. G., & Teece, D. J. (2010). Innovation spillovers and the "dirt road" fallacy: The intellectual bankruptcy of banning optional transactions for enhanced delivery over the internet. Journal of Competition Law and Economics, 6(3), 521–594. [perma.cc/G34S-UEU7]

attacks, and to improve the overall efficiency and capacity of their networks. Such allowances are often made through exceptions to a broad prohibition on operators' differential treatment of internet traffic. Traffic management practices that improve overall network performance, efficiency, or capacity without negatively impacting the quality of experience (QoE) for users should be allowed and encouraged. With appropriate safeguards, regulators may even wish to allow operators to improve the performance of one or more classes of network traffic, so long as the QoE for users of other classes of traffic or applications is not degraded. Often, these regulatory exceptions have assessed management practices according to whether they are "reasonable" or "non-commercial," but network slicing may add complexity to regulatory determinations about these and other approaches to network traffic management.

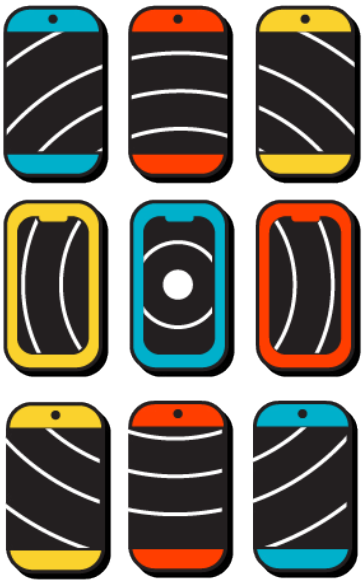

Network slicing is a powerful tool for managing network traffic, but its use as such should require regulatory oversight. Because network slicing allows operators to group similar kinds of traffic or endpoints together into virtual networks designed to provide optimal support, it could allow network operators to maximize the capacity and performance of their networks. However, it is no small task to sort and divide different kinds of network traffic so that each class of traffic can be given optimal treatment. It is even more difficult to do so without unduly favoring or disfavoring certain endpoints or applications. Further, it is not clear whether applications involving a blend of different service-level needs would benefit from this type of optimization. Adding to the complexity, technical constraints on the ability of network slices to perform within their service-level parameters may limit the number of endpoints a slice can serve, the volume of traffic it can carry, and other aspects of network capacity. Such limitations would require choosing which services or endpoints are included in a slice, potentially excluding others that would benefit from the same treatment. Thus, even without intent to discriminate, network operators may face difficulty in using network slices to equitably manage traffic on their networks.

One way operators could address the limitations imposed by technical constraints on network slices is to create multiple identical slices, as needed, to support demands in real-time. Indeed, the ability to quickly reconfigure network functions in response to changes in network demands is one of the major benefits of software-defined networking and network slicing. So if a network experiences a higher demand for a particular type of slice than a single slice can support, operators can reallocate underutilized network capacity toward additional slices. As long as network operators do not receive compensation or unfairly benefit their subsidiaries or affiliates through their management practices, the use of slices for traffic management should be allowed.

Ideally, network operators will use network slicing to improve the functionality of existing online applications and to preserve some capacity for the development of new network-based services reliant on guaranteed levels of service.

## Consumer Choice

User control is an overarching principle for individual rights and should guide the application of net neutrality regulations to specific new technologies, including network slicing. Where network slices are used for different quality of experience in internet usage, the choice of applications, services, and priorities for access must be made by the user.

As in the Body of European Regulators for Electronic Communications (BEREC) guidelines, these shouldn't be pre-selected configurations for the benefit of the network provider's business arrangements, but selections by the user of how they wish to prioritize their own needs and applications.[25] Producers of smartphone and desktop operating systems may play an important role in providing usable functionality to users in configuring this class of network choices.

And as we noted previously, where slicing or prioritization is made available as a consumer choice, consumer protection measures will be necessary, including transparency about speed, latency, pricing, and conditions.

---

# Recommendations

**B**ased on this technical assessment of network slicing, the potential impacts on important values (including competition, privacy, and neutrality), and possible regulatory approaches, we summarize recommendations for both operators and regulators to enable network slicing while supporting an open internet and human rights. Improving network efficiency in general, and spectral efficiency in particular, are worthy goals. But we urge regulators to scrutinize network operators' claims that regulations protecting the principles of net neutrality inhibit their ability to improve efficiency through network slicing and to take a careful approach toward the development of network slicing and preserving the open and flat nature of the internet.

### Monitoring, Reporting, and Transparency

It's early days for network slicing, and for the use of new technologies and standards including software-defined networking, mobile edge computing, virtualized network functions, and orchestration. Regulators should regularly ask for, and network providers should regularly contribute to, reporting on how network slicing is currently deployed and the potential impact of those deployments. Independent research, which may require funding or direct work from regulators, would be especially valuable in settling contentious questions.

### Additionality, Not Cannibalization

Network slicing is proposed as a way to efficiently use radio spectrum and new networking configurations to provide novel services that are otherwise infeasible: like gathering many sensor data readings in high-latency low-bandwidth low-power ways, or providing low-latency infrastructure for vehicle guidance. Providers should demonstrate, and regulators should require, that network slicing is only being used to add these new types of functionalities, to provide greater efficiencies, or to bring other benefits, rather than dividing up the internet into tiers and cannibalizing the general internet in order to distort the internet towards higher-rent specialization. Indeed, access to the internet must expand, and its quality should continue to be increased even as other services are added.

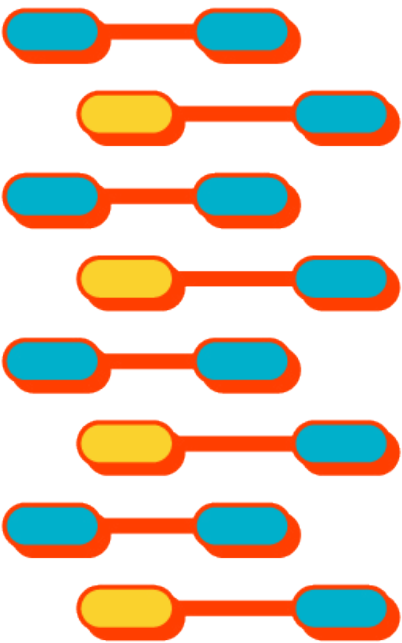

### Reasonable Network Management

Network slicing techniques may prove to be a powerful way to bring efficiency to 5G networks, including handling congestion in very crowded areas or providing nearby access to common resources. But network management needs oversight to guard against intentional or unintentional unreasonable discriminatory impacts on internet services.

### Open to All, Without Preferential Treatment

If net neutrality is going to provide equality of voice, access to network slicing needs to be provided openly, to all applications and endpoints, without preferential or anti-competitive treatment. This may require affirmative efforts to provide openness to orchestration arrangements of end-to-end network slicing or to standardization processes.

Attempts to closely tie services into the network itself might seem tempting for revenue generation, but they ultimately undermine the shared success of the internet and the end-to-end design principle. Configuration of application-specific network slices could also interfere with ongoing efforts to improve privacy protections and censorship circumvention.

As a regulatory matter, prohibitions on paid prioritization and self-preferencing must be maintained or re-enacted. Network slicing should be applied for general service and performance characteristics, rather than as a preference for individual applications.

### User Control

User control is fundamental and necessary, but not always sufficient, to support individual rights online. It may be challenging to describe meaningful functionality to users for these virtualized networks, but we invite implementation experience here and believe that the client endpoint is the preferred location for such control. Regulators should monitor these practices so that choices are clear and fair.

## Consumer Protection and Baseline Privacy

Telecommunications regulation must go hand-in-hand with consumer protection and enforcement of fundamental privacy rights. Particularly for subtle or complex networking functionality, there will be pricing and service implications that may not always be clear to customers, and access to personal data about sensitive network activities, where stronger protection is necessary.

—

**As a final recommendation, regulators, researchers, network operators, hardware vendors, consumer advocates, internet architecture experts, application providers, and users of the internet should continue to discuss and refine:** 1) how these network slicing technologies can be used; 2) what their impacts will be; and 3) how to support new network capabilities with even stronger protections, across jurisdictions, for competition, neutrality, privacy, and the long-term success of the internet.

This report cannot predict every application of network slicing techniques. Strong protections in regulation for net neutrality, competition, and privacy are necessary to protect consumers and enable a healthy competitive marketplace. But that should be only our baseline, not the end of our work: how the internet is designed, deployed, and maintained is also an ongoing technical and business challenge, and we want to take full advantage of the opportunities that new networking technology provides in addition to mitigating its risks. A careful and effective approach to stewardship of the internet requires legal protections for fundamental rights, but also regular on-the-ground technical analysis and multistakeholder discussion of architectural designs.

[cdt.org](cdt.org)

[cdt.org/contact](cdt.org/contact)

**Center for Democracy & Technology**
1401 K Street NW, Suite 200
Washington, D.C. 20005

+1 (202) 637-9800

[@CenDemTech](@CenDemTech)

[@CenDemTech@techpolicy.social](@CenDemTech@techpolicy.social)

CENTER FOR DEMOCRACY & TECHNOLOGY